\documentclass[onecolumn]{autart}
\usepackage[round]{natbib}
\usepackage{graphicx}
\usepackage{graphpap}
\usepackage{amsmath}
\usepackage{amsfonts}
\usepackage{graphicx}
\usepackage{amssymb}
\usepackage{algorithm, algpseudocode, algpascal}
\usepackage{color}

\definecolor{light}{gray}{.65}
\input{tcilatex}

\newtheorem{Def}{Definition}
\newtheorem{Rem}{Remark}
\newtheorem{Exp}{Example}
\begin{document}

\begin{frontmatter}

\title{A general representation form of system data with feedback control and fundamental
lemma as examples}

%\thanks[footnoteinfo]{Part of Sections 3-5 was presented at the 9th IFAC Symposium SAFEPROCESS held in Paris, France, September 2015.}

\author[Ding]{Steven X. Ding}  \ead{steven.ding@uni-due.de}  and
\author[linlin]{Linlin Li} \ead{linlin.li@ustb.edu.cn}

\address[Ding]{Institute for Automatic Control and Complex Systems, University of Duisburg-Essen, 47057 Duisburg, Germany}
\address[linlin]{School of Automation and Electrical Engineering, University of Science and Technology Beijing, 100083 Beijing, China}
                             % keyword list or with the
                                          % help of the Automatica
                                          % keyword wizard

\begin{abstract}                          % Abstract of not more than 200 words.
This note introduces a general data representation of
dynamic systems.
\end{abstract}

\end{frontmatter}

\section{Data representation}

Without loss of generality, we consider, in the sequel, discrete-time linear
time-invariant systems (LTI) modelled by%
\begin{equation}
y(z)=G(z)u(z),y\in \mathbb{C}^{p},u\in \mathbb{C}^{m}  \label{eq2-1}
\end{equation}%
with $u$ and $y$ as the plant input and output vectors. It is assumed that $%
G $ is a proper real-rational matrix and its minimal state space realisation
is given by 
\begin{align}
x(k+1)& =Ax(k)+Bu(k),x(0)=x_{0},  \label{eq2-2a} \\
y(k)& =Cx(k)+Du(k),  \label{eq2-2b}
\end{align}%
where $x\in \mathbb{R}^{n}$ is the state vector with the initial value $%
x_{0} $. Matrices $A,B,C,D$ are appropriately dimensioned real constant
matrices. Coprime factorisations of a transfer function matrix over $%
\mathcal{RH}_{\infty }$ give further system representation forms and
factorise a transfer function matrix into two $\mathcal{RH}_{\infty }$
transfer matrices \citep{Vidyasagar85}.

\begin{Rem}
Hereafter, \textnormal{the domain} variable $z$\ or $k$ may be \textnormal{%
dropped out, }when there is no risk of confusion.
\end{Rem}

The left and right coprime factorisations (LCF and RCF) of $G$ are given by 
\begin{equation}
G(z)=\hat{M}^{-1}(z)\hat{N}(z)=N(z)M^{-1}(z),  \label{eq2-3}
\end{equation}%
where $\left( \hat{M},\hat{N}\right) $ and $\left( M,N\right) $ are called
left and right coprime pairs (LCP and RCP), respectively. Corresponding to
them, there exist $\left( \hat{X},\hat{Y}\right) $ and $\left( X,Y\right) $
over $\mathcal{RH}_{\infty }$ so that 
\begin{equation}
\left[ 
\begin{array}{cc}
\hat{N} & \text{ \ }\hat{M}%
\end{array}%
\right] \left[ 
\begin{array}{c}
\hat{Y} \\ 
\hat{X}%
\end{array}%
\right] =\left[ 
\begin{array}{cc}
-\hat{N} & \text{ }\hat{M}%
\end{array}%
\right] \left[ 
\begin{array}{c}
-\hat{Y} \\ 
\hat{X}%
\end{array}%
\right] =I,\left[ 
\begin{array}{cc}
X & \text{ }Y%
\end{array}%
\right] \left[ 
\begin{array}{c}
M \\ 
N%
\end{array}%
\right] =I.  \label{eq2-5}
\end{equation}%
Hence, $\left( \hat{X},\hat{Y}\right) $ and $\left( X,Y\right) $ build RCP
and LCP as well. The state space representations of the coprime pairs $%
\left( \hat{M},\hat{N}\right) ,\left( M,N\right) ,\left( \hat{X},\hat{Y}%
\right) $ and $\left( X,Y\right) $ are given by%
\begin{align}
\hat{M}(z)& =\left( A-LC,-L,WC,W\right) ,\hat{N}(z)=\left(
A-LC,B-LD,WC,WD\right) ,  \label{eq2-4a} \\
M(z)& =\left( A+BF,BV,F,V\right) ,N(z)=\left( A+BF,BV,C+DF,DV\right) ,
\label{eq2-4b} \\
\hat{X}(z)& =\left( A+BF,LW^{-1},C+DF,W^{-1}\right) ,\hat{Y}(z)=\left(
A+BF,LW^{-1},-F,0\right) ,  \label{eq2-4c} \\
X(z)& =\left( A-LC,B-LD,-V^{-1}F,V^{-1}\right) ,Y(z)=\left(
A-LC,L,-V^{-1}F,0\right) .  \label{eq2-4d}
\end{align}%
They are parameterised by $\left( F,L,V,W\right) $ with Schur matrices $%
A_{F}=A+BF,A_{L}=A-LC$ and invertible matrices $\left( V,W\right) .$
Attributed to the above coprime factorisation relations, the well-known
double Bezout identity holds \citep{Vidyasagar85} 
\begin{equation}
\left[ 
\begin{array}{cc}
X(z) & \text{ }Y(z) \\ 
-\hat{N}(z) & \text{ }\hat{M}(z)%
\end{array}%
\right] \left[ 
\begin{array}{cc}
M(z) & \text{ }-\hat{Y}(z) \\ 
N(z) & \text{ }\hat{X}(z)%
\end{array}%
\right] =\left[ 
\begin{array}{cc}
I\text{ } & 0\text{ } \\ 
0\text{ } & I\text{ }%
\end{array}%
\right] .  \label{eq2-13}
\end{equation}%
A stable kernel representation of $G$ is 
\begin{equation}
r_{y}(z)=\hat{M}(z)y(z)-\hat{N}(z)u(z),  \label{eq2-6}
\end{equation}%
known as an observer-based residual generator \citep{Ding90}. Its dual form, 
\begin{equation}
\left[ 
\begin{array}{c}
u(z) \\ 
y(z)%
\end{array}%
\right] =\left[ 
\begin{array}{c}
M(z) \\ 
N(z)%
\end{array}%
\right] r_{u}(z),  \label{eq2-7}
\end{equation}%
is called stable image representation. Hereby, the input vector is defined
as $u(k)=F\hat{x}(k)+Vr_{u}(k),$ which can be \textit{interpreted} as an
observer-based state feedback with $r_{u}\in \mathbb{R}^{m}$ as a reference
signal. It turns out 
\begin{equation}
\left[ 
\begin{array}{c}
u \\ 
y%
\end{array}%
\right] =\left[ 
\begin{array}{cc}
M & \text{ }-\hat{Y} \\ 
N & \text{ }\hat{X}%
\end{array}%
\right] \left[ 
\begin{array}{cc}
X & \text{ }Y \\ 
-\hat{N} & \text{ }\hat{M}%
\end{array}%
\right] \left[ 
\begin{array}{c}
u \\ 
y%
\end{array}%
\right] =\left[ 
\begin{array}{c}
M \\ 
N%
\end{array}%
\right] r_{u}+\left[ 
\begin{array}{c}
-\hat{Y} \\ 
\hat{X}%
\end{array}%
\right] r_{y}.  \label{eq2-8}
\end{equation}%
A state space realisation of (\ref{eq2-8}) is given by 
\begin{gather}
\hat{x}(k+1)=A_{F}\hat{x}(k)+B_{F}r_{u}(k)+Lr_{0}(k),\hat{x}(0)=0,
\label{eq2-9} \\
\left[ 
\begin{array}{c}
u(k) \\ 
y(k)%
\end{array}%
\right] =\left[ 
\begin{array}{c}
F \\ 
C_{F}%
\end{array}%
\right] \hat{x}(k)+\left[ 
\begin{array}{c}
V \\ 
D_{F}V%
\end{array}%
\right] r_{u}(k)+\left[ 
\begin{array}{c}
0 \\ 
D%
\end{array}%
\right] r_{0}(k),  \label{eq2-9a} \\
r_{0}=y-\hat{y}=y-C\hat{x}-Du,B_{F}=BV,C_{F}=C+DF,D_{F}=VD,  \notag
\end{gather}%
whereas the state space realisations of (\ref{eq2-6}) and (\ref{eq2-7}) are
described by%
\begin{gather}
\hat{x}(k+1)=A_{L}\hat{x}(k)+B_{L}u(k)+Ly(k),\hat{x}(0)=0,  \label{eq2-10} \\
\left[ 
\begin{array}{c}
r_{u}(k) \\ 
r_{y}(k)%
\end{array}%
\right] =\left[ 
\begin{array}{c}
V^{-1}\left( u(k)-F\hat{x}(k)\right) \\ 
W\left( y(k)-\hat{y}(k)\right)%
\end{array}%
\right] =\left[ 
\begin{array}{c}
-V^{-1}F \\ 
C_{L}%
\end{array}%
\right] \hat{x}(k)+\left[ 
\begin{array}{c}
V^{-1} \\ 
D_{L}%
\end{array}%
\right] u(k)+\left[ 
\begin{array}{c}
0 \\ 
W%
\end{array}%
\right] y(k),  \label{eq2-10a} \\
B_{L}=B-LD,D_{L}=-WD,C_{L}=-WC.  \notag
\end{gather}%
Below, the control-theoretic interpretations of (\ref{eq2-9})-(\ref{eq2-9a})
and (\ref{eq2-10})-(\ref{eq2-10a}) are shortly explained. Residual $r_{y}$
represents uncertainties is a well-known fact widely applied in fault
diagnosis research \citep{Ding2008}. It describes influences of any types of
uncertainties in the plant (\ref{eq2-1}) on $y$ \citep{Ding2020}. System (\ref%
{eq2-9}) is the system response (i) to the input that is implicitly subject
to $u=F\hat{x}+Vr_{u}$ for some $\left( F,V,r_{u}\right) $ and (ii) to the
uncertainties $r_{y}$ in the system. Accordingly, the data representation (%
\ref{eq2-8}) is understood as the process input-output data $\left(
u,y\right) $ comprises two parts representing the nominal dynamics $\left[ 
\begin{array}{c}
M \\ 
N%
\end{array}%
\right] r_{u}$ and uncertainty dynamics $\left[ 
\begin{array}{c}
-\hat{Y} \\ 
\hat{X}%
\end{array}%
\right] r_{y},$ respectively.

For practical applications, systems (\ref{eq2-9})-(\ref{eq2-9a}) and (\ref%
{eq2-10})-(\ref{eq2-10a}) are of outstanding interests. While system (\ref%
{eq2-9})-(\ref{eq2-9a}) fully depicts the system behaviour under all
possible operation conditions and thus can be applied for serving as a
real-time digital twin towards real-time simulation and process monitoring,
system (\ref{eq2-10})-(\ref{eq2-10a}) reflects the variations in the plant
and the controller. As residual generators and detectors they are useful for
fault and attack detection. In particular, it is to emphasise that a single
state observer is at the core of both systems. That implies, online
implementation of systems (\ref{eq2-9})-(\ref{eq2-9a}) and (\ref{eq2-10})-(%
\ref{eq2-10a}) merely claims computations of a $n$-the order difference
equation.

To conclude the above discussion, the following definitions are introduced.

\begin{Def}
The representation (\ref{eq2-8}) is called general data representation.
\end{Def}

On account of the general data representation (\ref{eq2-8}), the concepts of
image/kernel and residual subspaces are introduced \citep{Ding2026}.

\begin{Def}
Given system $G,$ the image subspace $\mathcal{I}_{G}$ or equivalently the
kernel subspace $\mathcal{K}_{G}$, and residual subspace $\mathcal{R}_{G}$
are respectively defined by 
\begin{align}
\mathcal{K}_{G}& =\left\{ \left[ 
\begin{array}{c}
u \\ 
y%
\end{array}%
\right] :\left[ 
\begin{array}{cc}
-\hat{N} & \text{ \ }\hat{M}%
\end{array}%
\right] \left[ 
\begin{array}{c}
u \\ 
y%
\end{array}%
\right] =0,\left[ 
\begin{array}{c}
u \\ 
y%
\end{array}%
\right] \in \mathcal{H}\right\} \subset \mathcal{H}, \\
\mathcal{I}_{G}& =\left\{ \left[ 
\begin{array}{c}
u \\ 
y%
\end{array}%
\right] :\left[ 
\begin{array}{c}
u \\ 
y%
\end{array}%
\right] =\left[ 
\begin{array}{c}
M \\ 
N%
\end{array}%
\right] r_{u},r_{u}\in \mathcal{H}\right\} \subset \mathcal{H}, \\
\mathcal{R}_{G}& =\left\{ \left[ 
\begin{array}{c}
u \\ 
y%
\end{array}%
\right] :\left[ 
\begin{array}{c}
u \\ 
y%
\end{array}%
\right] =\left[ 
\begin{array}{c}
-\hat{Y} \\ 
\hat{X}%
\end{array}%
\right] r_{y},r_{y}\in \mathcal{H}\right\} \subset \mathcal{H},
\end{align}%
where $\mathcal{H}$ denotes the Hilbert space under consideration.
\end{Def}

Accordingly, the general data representation (\ref{eq2-8}) can be expressed
as a decomposition of $\left( u,y\right) $ in the image and residual
subspaces. The subspaces $\mathcal{I}_{G}$ and $\mathcal{R}_{G}$ are,
attributed to the Bezout identity (\ref{eq2-13}), two complementary
subspaces, i.e.%
\begin{equation}
\mathcal{H}=\mathcal{I}_{G}\oplus \mathcal{R}_{G}.
\end{equation}

To illustrate application of the general data representation (\ref{eq2-8}),
the subsequent examples are given.

\begin{Exp}
It is noteworthy that the data representation (\ref{eq2-8}) holds for both
open and closed-loop system configurations. The latter is of significant
interest in control theory and engineering. In the context of the data
representation (\ref{eq2-8}), a feedback controller is nothing other than a
setting of $\left( F,L,V,W\right) $ and task-dependent $r_{u}.$
Specifically, for an observer-based state feedback system, it is $u=F\hat{x}%
+Vv$ with feedback gain, pre-filter and observer gain $\left( F,L,V,W\right) 
$ as well as $r_{u}=v$ as a reference signal. The same is true with a
stabilising feedback controller $u=Ky+\bar{v},$ incl. a PI-controller, which
can be knowingly described by 
\begin{equation}
K=-\left( \hat{Y}-MQ\right) \left( \hat{X}+NQ\right) ^{-1}=-\left( X+Q\hat{N}%
\right) ^{-1}\left( Y-Q\hat{M}\right)
\end{equation}%
with the coprime pairs $\left( \hat{M},\hat{N}\right) ,\left( M,N\right)
,\left( \hat{X},\hat{Y}\right) $ and $\left( X,Y\right) $ given in (\ref%
{eq2-4a})-(\ref{eq2-4d}) and the parametrisation system $Q\in \mathcal{RH}%
_{\infty }$ \citep{Youla1976b}. Attributed to the extended form of Bezout
identity (\ref{eq2-13}), 
\begin{equation}
\left[ 
\begin{array}{cc}
M & \text{ }-\hat{Y}+MQ \\ 
N & \text{ }\hat{X}+NQ%
\end{array}%
\right] \left[ 
\begin{array}{cc}
X+Q\hat{N} & \text{ }Y-Q\hat{M} \\ 
-\hat{N} & \text{ }\hat{M}%
\end{array}%
\right] =\left[ 
\begin{array}{cc}
I\text{ } & 0\text{ } \\ 
0\text{ } & I\text{ }%
\end{array}%
\right] ,
\end{equation}%
$u$ is equivalent to an observer-based state feedback controller \citep%
{Ding2020} 
\begin{equation}
u=F\hat{x}+Qr_{y}+v,v=\left( X+Q\hat{N}\right) \bar{v}.  \label{eq2-11}
\end{equation}%
It is worth emphasising that $F\hat{x}+Qr_{y}$ is an observer for $Fx$ \citep%
{DGF94,Ding2026}. As a result, the settings are $\left( F,L,V,W\right) $ and 
$r_{u}=Qr_{y}+v.$ It is obvious that the system response $\left( u,y\right) $
is 
\begin{equation}
\left[ 
\begin{array}{c}
u \\ 
y%
\end{array}%
\right] =\left[ 
\begin{array}{c}
M \\ 
N%
\end{array}%
\right] r_{u}+\left[ 
\begin{array}{c}
-\hat{Y} \\ 
\hat{X}%
\end{array}%
\right] r_{y}=\left[ 
\begin{array}{c}
M \\ 
N%
\end{array}%
\right] v+\left[ 
\begin{array}{c}
\text{ }-\hat{Y}+MQ \\ 
\text{ }\hat{X}+NQ%
\end{array}%
\right] r_{y},  \label{eq2-12}
\end{equation}%
i.e. the response trajectory to the reference signal and the response to the
uncertainty (including any type of faults). It can be interpreted as an
approximation of the target system dynamics $\left[ 
\begin{array}{c}
M \\ 
N%
\end{array}%
\right] r_{u}$ by an observer-based, residual-driven system. The
aforementioned results illustrate a fundamental conclusion that a feedback
control is at its core an observer serving feedback of uncertainties in the
plant, represented by $r_{y}.$ This is an information (in the context of
uncertainties) aspect of feedback control, and one of the core principles of
the unified framework of control and detection \citep{Ding2026}, that
characterises feedback loop and data dynamics.
\end{Exp}

\begin{Exp}
Let $\left( u_{s}(k),y_{s}(k)\right) $ be $s$-samples input-output data of $%
G,$ 
\begin{equation*}
\begin{bmatrix}
\,u_{s}(k)\, \\ 
\,y_{s}(k)\,%
\end{bmatrix}%
=\left[ 
\begin{array}{c}
u(k+1) \\ 
\vdots \\ 
u(k+s) \\ 
y(k+1) \\ 
\vdots \\ 
y(k+s)%
\end{array}%
\right] \in \mathbb{R}^{s\left( m+p\right) },s\geq n,
\end{equation*}%
a notation widely adopted in data-driven parity space study \citep{DZNDH2009}%
. Consider the nominal dynamics, i.e. $r_{y}=0.$ On account of (\ref{eq2-8}%
), the $s$-samples input-output data of $G$ is given by%
\begin{gather}
\begin{bmatrix}
\,u_{s}(k)\, \\ 
\,y_{s}(k)\,%
\end{bmatrix}%
=I_{G,s}r_{u,s+n}(k-n),r_{u,s+n}(k-n)=\left[ 
\begin{array}{c}
r_{u}(k-n+1) \\ 
\vdots \\ 
r_{u}(k+s)%
\end{array}%
\right] \in \mathbb{R}^{p\left( s+n\right) },  \label{eq2-15} \\
I_{G,s}=\left[ 
\begin{array}{c}
M_{s} \\ 
N_{s}%
\end{array}%
\right] ,\left[ 
\begin{array}{c}
M_{s} \\ 
N_{s}%
\end{array}%
\right] =\left[ 
\begin{array}{c}
\mathcal{T}_{s,s+n}\left( G^{M}\right) \\ 
\mathcal{T}_{s,s+n}\left( G^{N}\right)%
\end{array}%
\right] ,  \notag \\
G_{l}^{M}=\left\{ 
\begin{array}{l}
FA_{F}^{n+l-1}B,l\geq 1 \\ 
I,l=0 \\ 
0,l<0,%
\end{array}%
\right. \mathcal{T}_{s,s+n}\left( G^{M}\right) _{i,j}=G_{i-j}^{M},  \notag \\
G_{l}^{N}=\left\{ 
\begin{array}{l}
FA_{F}^{n-1+l}B,l\geq 1 \\ 
D,l=0 \\ 
0,l<0,%
\end{array}%
\right. \mathcal{T}_{s,s+n}\left( G^{N}\right) _{i,j}=G_{i-j}^{N}  \notag
\end{gather}%
with the Toeplitz matrices $\mathcal{T}_{s,s+n}\left( G^{M}\right) $ and $%
\mathcal{T}_{s,s+n}\left( G^{N}\right) ,$ and $i,j,l\in \mathbb{Z}$ \citep%
{Ding2026,LDZ2026}. Attributed to the rank condition, 
\begin{equation}
rank\left[ 
\begin{array}{c}
M_{s} \\ 
N_{s}%
\end{array}%
\right] =sm+n,  \label{eq2-16}
\end{equation}%
the s-samples image representation $\left( M_{s},N_{s}\right) $ can be
realised using input-output data sets expressed by Hankel matrices of $%
u_{s}(k)$ and $y_{s}(k),$ respectively, 
\begin{gather}
\left[ 
\begin{array}{c}
M_{s} \\ 
N_{s}%
\end{array}%
\right] =\left[ 
\begin{array}{c}
\mathcal{H}_{s}(u_{[k+1:k+N]})\hspace{-2pt} \\ 
\mathcal{H}_{s}(y_{[k+1:k+N]})\hspace{-2pt}%
\end{array}%
\right] \Longrightarrow 
\begin{bmatrix}
\,u_{s}(k)\, \\ 
\,y_{s}(k)\,%
\end{bmatrix}%
=\left[ 
\begin{array}{c}
\mathcal{H}_{s}(u_{[k+1:k+N]})\hspace{-2pt} \\ 
\mathcal{H}_{s}(y_{[k+1:k+N]})\hspace{-2pt}%
\end{array}%
\right] r_{u,s+n}(k-n),  \label{eq2-18} \\
\mathcal{H}_{s}(u_{[k+1:k+N]})\hspace{-2pt}=\hspace{-2pt}\left[ \hspace{-2pt}%
\begin{array}{ccc}
u_{s}(k) & \cdots & \hspace{-2pt}u_{s}(k\hspace{-2pt}+\hspace{-2pt}N\hspace{%
-2pt}-\hspace{-2pt}s)%
\end{array}%
\hspace{-2pt}\right] \hspace{-4pt}\in \hspace{-2pt}\mathbb{R}^{sm\times
\left( s+n\right) m},  \notag \\
\mathcal{H}_{s}(y_{[k+1:k+N]})\hspace{-2pt}=\hspace{-2pt}\left[ \hspace{-2pt}%
\begin{array}{ccc}
y_{s}(k) & \cdots & \hspace{-2pt}y_{s}(k\hspace{-2pt}+\hspace{-2pt}N\hspace{%
-2pt}-\hspace{-2pt}s)%
\end{array}%
\hspace{-2pt}\right] \hspace{-4pt}\in \hspace{-2pt}\mathbb{R}^{sp\times
\left( s+n\right) m},  \notag
\end{gather}%
for $N=\left( s+n\right) m+s-1.$ For practical applications, $N>>\left(
s+n\right) m+s-1.$ Hereby, the input Hankel matrix should satisfy 
\begin{equation}
rank\left( \mathcal{H}_{s}(u_{[k+1:k+N]})\right) =\left( s+n\right) m.
\label{eq2-17}
\end{equation}%
The expression (\ref{eq2-18}) is the so-called fundamental lemma \citep%
{WILLEMS2005}. The rank condition (\ref{eq2-17}) is the requirement of
persistently exciting of order $\left( s+n\right) m$. For the proving
details, the reader is referred to \citep{Ding2026,LDZ2026}. In the framework
of behavioural theory \citep{Willems1998}, the widely celebrated fundamental
lemma \citep{WILLEMS2005} has received enormous attention in the recent
decade with successful applications to data-driven MPC \citep{MARKOVSKY2021}.
\end{Exp}

Further useful applications of the data representation (\ref{eq2-8}) are,
for instance, safe multi-agent systems \citep{Ding2026}, and the recent study
on safe and secure cyber-physical control systems (CPCSs) \citep%
{Ding2026,LD2026}, which considerably simplifies system analysis and
synthesis, and can be applied to exploring more complicated problem
settings. It is observed that the publication \citep{DLautomatica2022} under
this aspect did not receive nameable resonance.

\section{Conclusion}

This note has described the general data representation. For continuous-time
LTI, LTV, and polynomial systems as well as nonlinear systems under certain
conditions, the general data representation exists \citep{Ding2026}.

%\bibliographystyle{model5-names}
%\bibliography{ieeepesp}

\end{document}